\documentclass[aps,prd,onecolumn,groupedaddress,showpacs,nofootinbib,amssymb]{revtex4}
\usepackage[dvips]{graphicx}
\usepackage{amssymb}
\usepackage{amsmath}
\usepackage{graphicx,,color}
\usepackage{amsfonts}
\usepackage{bm}
\usepackage{cancel}
\usepackage{comment}

\graphicspath{ {./Images/} }

\newcommand\be{\begin{equation}}
\newcommand\ee{\end{equation}}

\allowdisplaybreaks[4]

\begin{document}

\title{Canonical Scalar Field Inflation with a Woods-Saxon Potential}
\author{V.K.~Oikonomou,$^{1,2,3,4}$\,\thanks{v.k.oikonomou1979@gmail.com}\,\thanks{nchatzar@physics.auth.gr}, N.Th.~Chatzarakis$^{1}$}
\affiliation{
$^{1)}$ Department of Physics, Aristotle University of Thessaloniki, Thessaloniki 54124, Greece\\
$^{2)}$ Laboratory for Theoretical Cosmology, Tomsk State University
of Control Systems
and Radioelectronics, 634050 Tomsk, Russia (TUSUR)\\
$^{3)}$ Tomsk State Pedagogical University, 634061 Tomsk, Russia\\
$^{4)}$ Theoretical Astrophysics, IAAT, University of T\"{u}bingen, Germany
}

\tolerance=5000

\begin{abstract}
This paper focuses on the realization of an inflationary model from a canonical scalar field theory with a Woods-Saxon potential, in the slow-roll approximation. Our analysis indicates that the observable quantities derived theoretically from our model, namely the spectral index of the primordial scalar curvature perturbations and the tensor-to-scalar ratio, are compatible with the latest Planck collaboration data. We also discuss the qualitative features of the potential, and we show that the value of the scalar field for which the graceful exit occurs, coincides with the inflection point of the scalar potential. We also attempt to study the post-inflation reheating phase of the model, in order to further examine the viability of the Woods-Saxon scalar field model, and as we demonstrate the results indicate viability of the model for this era too, however the instantaneous reheating is not allowed for the model at hand.
\end{abstract}


\maketitle

\section{Introduction}

The primordial era of the Universe after it exited the quantum era, is undoubtedly one mysterious period of evolution, and at the same time perhaps the only link of classical physics to the post-quantum Universe. Two scenarios seem for the moment to consistently describe the primordial evolution, either the inflationary scenario \cite{Linde:2007fr,Gorbunov:2011zzc,Lyth:1998xn,Martin:2018ycu} or the bouncing cosmology scenario \cite{reviews1,Brandenberger:2012zb,Brandenberger:2016vhg,Battefeld:2014uga,Novello:2008ra,Cai:2014bea,deHaro:2015wda,Lehners:2011kr,Lehners:2008vx, Cheung:2016wik,Cai:2016hea}. Both of them have become increasingly popular, as they appear to resolve the majority of problems the standard Big Bang cosmology faced, with the main attribute of both being that these are able to generate a nearly scale invariant power spectrum of primordial curvature perturbations. Every theoretical model eventually is confronted with the observational data coming from the latest Planck collaboration \cite{Akrami:2018odb}, and many models of modified gravity and standard scalar-tensor gravity, manage still to successfully be consistent with observational constraints, see for example the reviews Refs. \cite{reviews1,reviews2,reviews3,reviews4,reviews5,reviews6}. Such an example are the $f(R)$ models of gravity, that can provide a viable cosmological evolution for a series of completely different inflationary scenarios \cite{Nojiri:2003ft,Starobinsky:1980te,Nojiri:2006gh,Capozziello:2006dj,Capozziello:2005ku,Carloni:2004kp,Noh:2001ia,Barrow:1988xh,Liddle:1994dx,Hwang:2001qk,Hwang:2001pu,Hwang:1990re,Ferraro:2006jd, Nojiri:2007cq, Huang:2013hsb, Hwang:1995bv, Artymowski:2014gea,Amin:2015lnh, Brooker:2016oqa, Sebastiani:2015kfa, Odintsov:2016plw, Oikonomou:2015qfh, Odintsov:2015gba, Nojiri:2017qvx}, or for bouncing cosmology scenarios \cite{Odintsov:2015zza, Odintsov:2015ynk}.

Apart from the modified gravity description, the most traditional approach to describe the inflationary era is to use a slow-rolling canonical scalar field \cite{Linde:1981mu,Linde:1983gd,Lyth:1998xn,Linde:1993cn}.  Such formulations are very well-known in the literature, as their scope is to identify the form of the potential that can generate a viable inflationary scenario, a graceful exit from it, and eventually to provide a reheating phase right afterwards. In this sense, we may come closer to identifying the scalar field expected to have actually generated the early-time accelerating expansion, since both direct observations have so far been impossible, as well as the fundamental physics of so high energies remain unfolded. Several potentials have been utilized so far, in context of the minimal and non-minimal scalar coupled theories \cite{reviews1}.

In this paper we shall investigate in detail the phenomenological implications on the inflationary era, of a newly introduced Wood-Saxon potential, in the slow-roll approximation. This form of potential is known from nuclear physics studies \cite{Zhou:2003jv}, so in this work we shall assume that the scalar field potential has a Woods-Saxon form. We shall calculate the spectral index of the primordial scalar curvature perturbations and the tensor-to-scalar ratio and eventually we shall confront the theoretical predictions of the model with the latest Planck data. As we demonstrate the resulting phenomenology is compatible with the latest Planck data for a wide range of the free parameters values. In addition, we shall study the reheating era produced by the Woods-Saxon canonical scalar model, and we shall calculate in detail the duration of the reheating era and the reheating temperature. As we shall demonstrate, the results are within the allowed observational and theoretical predictions, however the instantaneous reheating process cannot be realized by the Woods-Saxon canonical scalar field model, so the maximum reheating temperature for the model at hand is not reached right after inflation. Nevertheless, a consistent reheating era can be produced by the Woods-Saxon scalar theory, for the same range of values of the free parameters that can generate a viable inflationary era.

The paper is organized as follows: In section II, we will provide an overview of the inflationary dynamics of a slow-rolling canonical scalar field and we introduce the Woods-Saxon scalar potential. We also calculate in detail the slow-roll indices and the inflationary indices, and accordingly we compare the results with the observational data coming form the Planck collaboration. In section IV, we study the post-inflationary reheating era and again compare our results to the observations. Finally, the conclusions follow at the end of the paper.

\section{Inflationary Dynamics: The Woods-Saxon Potential Case}

In the following, we always deal with a flat Friedmann-Robertson-Walker (FRW) Universe, with line element,
\begin{equation}\label{eq:FriedMetric}
ds^2 = -dt^2 + a(t)^2 \sum_i dx_i^{2} \, ,
\end{equation}
where $i=1,2,3$ denote the Euclidean spatial dimensions, and $a(t)$ is the scale factor, implying an expansion (or contraction) of the 3-dimensional spatial spacelike hypersurface over time. We also assume that the action of our theory has the form
\begin{equation}\label{eq:Action}
S = \int d^{4}x \sqrt{-g} \Big( \dfrac{1}{2\kappa^2} R - \dfrac{1}{2} g^{\mu\nu} \partial_{\mu}\phi \partial_{\nu}\phi - V(\phi) \Big) \, ,
\end{equation}
where $R$ is the Ricci scalar, $g$ is the determinant of the metric $g^{\mu\nu}$, $\phi = \phi(x^{\mu})$ is the scalar field with potential $V(\phi)$, and $\kappa^2 = 8\pi G $ in natural units. By varying the action with respect to the metric, we obtain,
\begin{equation}\label{eq:Friedmann}
H^2 = \dfrac{\kappa}{3} \rho_{(\phi)} \, ,
\end{equation}
and the Landau-Raychaudhuri equation
\begin{equation}\label{eq:Raychaudhuri}
\dot{H} = -\dfrac{\kappa}{2} \left( \rho_{(\phi)} + P_{(\phi)} \right) \, ,
\end{equation}
where $H = \dfrac{\dot{a}}{a}$ the Hubble rate, $\rho_{(\phi)} = - T_{00}^{(\phi)}$ the effective energy density and $P_{(\phi)} = T_{ij}^{(\phi)}$ the effective Pressure of the scalar field -where $T_{\mu\nu}^{(\phi)} = \dfrac{\delta \mathcal{L}_{\phi}}{\delta g^{\mu\nu}}$ the effective energy-momentum tensor. Specifically, the energy density and the pressure of the scalar field are,
\begin{equation}
\rho_{(\phi)} = \dfrac{1}{2} \dot{\phi} + V(\phi) \;\;\; \text{and} \;\;\; P_{(\phi)} = \dfrac{1}{2} \dot{\phi} - V(\phi) \, .
\end{equation}
In order for the system to be complete one needs a continuity equation and an effective equation of state for the scalar field. The first of these is easily obtained, either by varying the action of Eq. (\ref{eq:Action}) with respect to the scalar field, $\dfrac{\delta S}{\delta \phi} = 0$, or by assuming the covariant constancy of the effective energy-momentum tensor, $\nabla_{\mu} T^{\mu\nu} = 0$. Either way will eventually produce the equation of motion for the scalar field,
\begin{equation}\label{eq:KleinGordon}
\ddot{\phi} + 3 H \dot{\phi} + \frac{\partial V}{\partial \phi} = 0 \, .
\end{equation}
As for the effective equation of state, linking pressure and energy-density of the scalar field, we can simply assume that the effective fluid corresponding to the field is a perfect fluid, hence its barotropic index is easily found to be the quotient of the pressure over the energy density,
\begin{equation}\label{eq:EffEoS}
w_{eff} = \dfrac{P_{(\phi)}}{\rho_{(\phi)}} = \dfrac{\frac{1}{2} \dot{\phi} - V(\phi)}{\frac{1}{2} \dot{\phi} + V(\phi)} \, .
\end{equation}
In order for this scalar field to generate and inflationary model, we need a slow-rolling condition, in other words, we need the scalar field to slow-roll on its potential for sufficient time so long as for the scale factor of the Universe to increase sufficiently for the Cauchy, the horizon and the flatness problems to be resolved. This means that eventually the following condition needs to hold true,
\begin{equation}\label{eq:SlowRollCond}
V(\phi) \gg \dfrac{1}{2}\dot{\phi}^2 \, ,
\end{equation}
This simply means that the kinetic term must become really small (approximately zero) in comparison to the potential. In order to quantify this, we introduce the slow-roll indices, which in the case of General Relativity with minimally coupled scalar fields take the form,
\begin{equation}\label{eq:SlowRollInd1}
\epsilon = -\dfrac{\dot{H}}{H^2} \;\;\; \text{and} \;\;\; \eta = -\dfrac{\ddot{H}}{2 H \dot{H}} \, ,
\end{equation}
that measure the velocity and the acceleration of the Hubble rate. Furthermore, we can introduce the deceleration parameter, $q = -\Big( 1 + \dfrac{\dot{H}}{H^2} \Big)$, that is strongly related to the slow-roll indices,
\begin{equation}\label{eq:Deceleration}
q = \epsilon - 1 \;\;\; \text{and} \;\;\; \dot{q} = 2\epsilon \left( \epsilon - \eta \right) \, .
\end{equation}
In what we know, by imposing the condition (\ref{eq:SlowRollCond}) means that the Hubble rate must be proportional to the square of the potential and its derivative approximately zero, as can be seen from Eqs. (\ref{eq:Friedmann}), (\ref{eq:Raychaudhuri}) and (\ref{eq:EffEoS}),
\begin{equation}
H^2 \simeq \dfrac{\kappa}{3} V(\phi) \;\;\; \text{and} \;\;\; \dot{H} \simeq 0 \, ,
\end{equation}
which implies an approximate constancy for the Hubble rate. The slow-roll indices must take the form,
\begin{equation}\label{eq:SlowRollInc2}
\epsilon = \dfrac{3 \dot{\phi}}{2 V(\phi)} \simeq \dfrac{1}{2\kappa^2} \left( \dfrac{V'(\phi)}{V(\phi)} \right)^2 \;\;\; \text{and} \;\;\; \eta \simeq \dfrac{3 \dot{\phi}}{V(\phi)} \simeq \dfrac{1}{\kappa^2} \dfrac{ \left| V''(\phi) \right|}{V(\phi)} \, ,
\end{equation}
where the prime indicates differentiation with respect to the scalar field.  From Eq. (\ref{eq:SlowRollInc2}) we have that $\epsilon \simeq 2 \eta$ and that both of them must be essentially small ($\epsilon, \eta \ll 1$), as a consequence of the slow-roll condition. Introducing these to Eqs. (\ref{eq:EffEoS}) and (\ref{eq:Deceleration}), we observe that,
\begin{equation*}
w_{eff} \simeq -1 \;\;\; \text{and} \;\;\; q \simeq -1 \, ,
\end{equation*}
which means that, firstly the effective fluid corresponding the scalar field shares the same characteristics as the quantum vacuum energy, and secondly, that such an effective fluid allows for a highly accelerating expansion.

Due to the slow-roll condition and an imposed linearization, we can derive the spectral index of primordial scalar curvature perturbations, $n_{S}$, and the tensor-to-scalar ratio, $r$, as functions of the slow-roll indices,
\begin{align}
n_{S} &\simeq 1 - 6\epsilon + 2\eta \label{eq:SpectralInd} \, , \\
r &= 16\epsilon \label{eq:TensorScalarRatio} \, .
\end{align}
If the Universe is to gracefully exit the inflationary era and continue with the subsequent phases of evolution, then the slow-roll condition must no longer hold true. In that case, the kinetic term would again become comparable to the potential and the slow-roll indices values would increase. Especially $\epsilon$ must reach unity, which is exactly the condition for the graceful exit to occur. This means that the potential will reach some minimum and that the scale factor of the Universe has become sufficiently large. The reflection of these on the spectral index and the tensor-to-scalar ratio would be that they would move towards constant values of very specific value. Particularly, the spectral index must approach (but not reach) unity, while the tensor-to-scalar ration must become extremely small (certainly below $0.1$).

Observations of the Cosmic Microwave Background obtained by the Planck collaboration \cite{Akrami:2018odb}, indicate the following constraints on the values the spectral index and the tensor-to-scalar ratio,
\begin{equation}\label{eq:Observables}
n_{S} = 0.9649 \pm 0.042 \;\;\; \text{and} \;\;\; r < 0.064 \, .
\end{equation}
Based on these values, we can estimate the viability of each proposed model for the early-time expansion of the Universe.

\subsection{Inflationary Dynamics of the Woods-Saxon scalar field}

The inflationary scenario we propose is generated by a scalar field with a Woods-Saxon potential. This potential is defined as,
\begin{equation}\label{eq:WoodsSaxon}
V(\phi) = \dfrac{V_{0}}{1 + \beta e^{-\alpha\kappa\phi}} \, ,
\end{equation}
where $V_{0}$, $\alpha$, and $\kappa$ are real constants  with dimensions eV$^4$, eV and eV$^{-1}$, while $\beta$ is a real dimensionless constant. Generally $V_{0}$ represents the depth of the potential and $\alpha$ and $\beta$ are left as free parameters of the model.

\begin{figure}[h!] 
\centering
    \includegraphics[width=18pc]{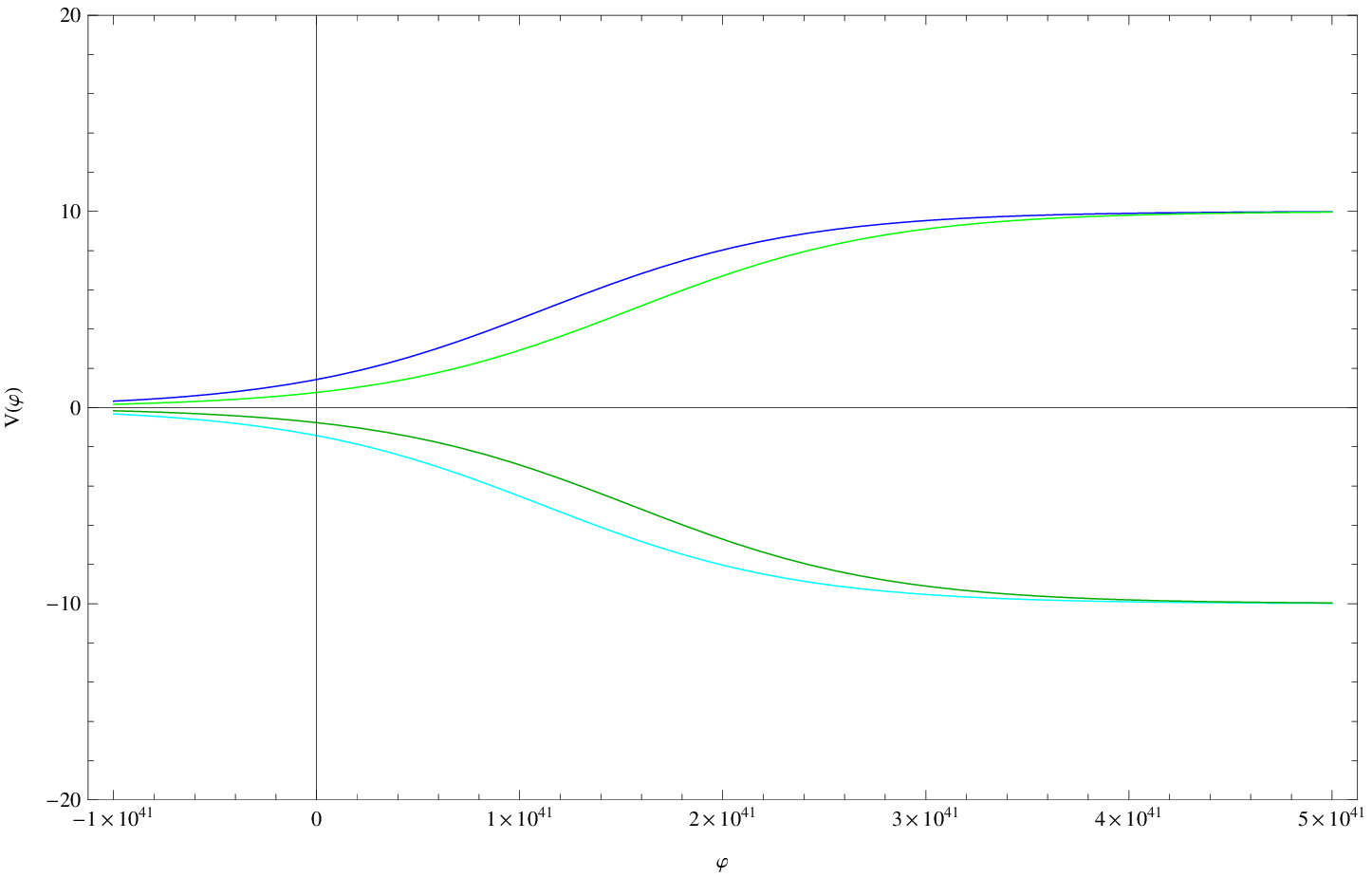}
    \includegraphics[width=18pc]{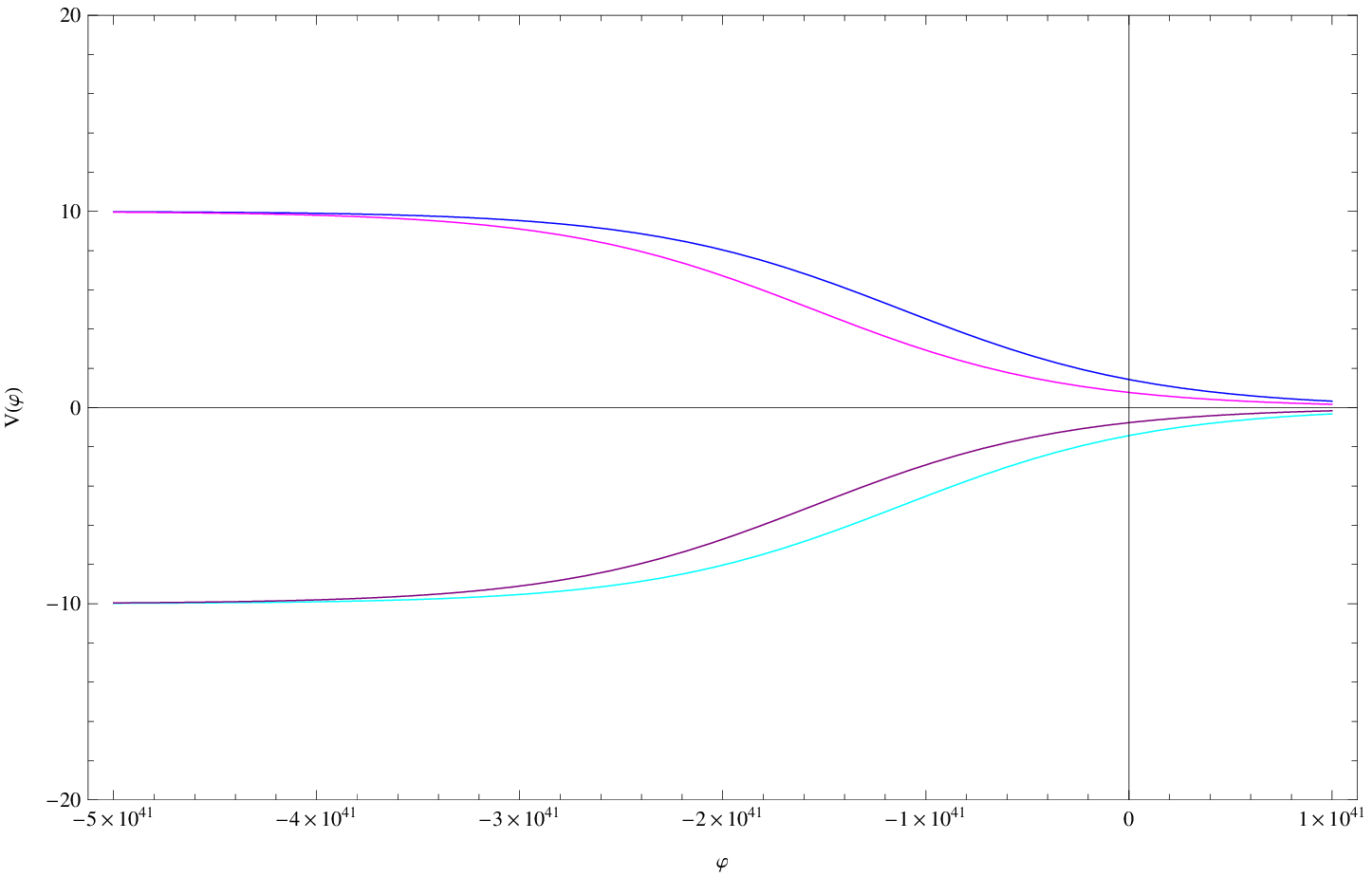}
\caption{ The form of the potential for $\kappa = 3.90584 \times 10^{-28}$eV$^{-1}$ and different values for the other parameters. In the left-hand-side $\alpha = 16$ and in the right-hand-side $\alpha = -16$; $V_{0}$ is either $10\,$eV$^{4}$ or $-10\,$eV$^{4}$, depending on whether the potential goes over or falls below zero; finally, blue and cyan curves stand for $\beta = 6$ and $\beta = 12$ respectively.}
\label{fig:Potential}
\end{figure}

It is obvious that the potential has no maxima or minima, but it has an inflection point located at
\begin{equation*}
\phi^{*} = \dfrac{\ln \beta}{\alpha \kappa} \, ,
\end{equation*}
which may not exist for $\beta < 0$ or for $\alpha = 0$; the inflection point lies in the negative if $0 < \beta < 1$ and $\alpha > 0$ or $\beta > 1$ and $\alpha < 0$, and lies in the positive in the opposite, while it is exactly $\phi^{*} = 0$ when $\beta = 1$. From the plots of the potential in Fig. \ref{fig:Potential}, we can see that $V_{0}$ is indeed the depth of the potential, $\alpha$ indicates whether the attainment of the depth will be for negative or positive values of the scalar field and, finally, $\beta$ denotes the ``stickiness'' of the potential, in other words, the speed of the convergence towards $V_{0}$. Just for the record, if $\beta < 0$, then the potential does not have its double plateau, but it rises to infinity from one side.

The presence of a plateau that reaches some edge and then drops, is a feature we esteem, since it can generate the slow-roll condition during inflation, but also the graceful exit after that. From Eq. (\ref{eq:SlowRollInc2}), we may calculate the slow-roll indices for this scalar field during inflation, which are,
\begin{equation}\label{eq:SlowRoll}
\epsilon = \dfrac{\alpha ^2 \beta ^2 e^{-2 \alpha \kappa \phi}}{2 \left( \beta e^{-\alpha \kappa  \phi} + 1\right)^2} \;\;\; \text{and} \;\;\; \eta = -\dfrac{\alpha ^2 \beta \left(e^{\alpha \kappa \phi}-\beta \right)}{\left( e^{\alpha \kappa \varphi } + \beta \right)^2} \, .
\end{equation}
In order for the inflationary era to reach an end, we shall assume that $\epsilon=1$ and solve with respect to the scalar field, so we obtain,
\begin{equation}
\phi_{fin} = \dfrac{\ln \left( -\frac{\beta}{2} \left( 2 + \sqrt{2}\alpha \right) \right)}{\alpha\kappa } \, ,
\end{equation}
which is the value of the scalar field as the Universe exits inflation. Interestingly enough, this value coincides with the inflection point for $\alpha = \dfrac{\sqrt{2}}{2}$.

\begin{figure}[h!] 
\centering
    \includegraphics[width=18pc]{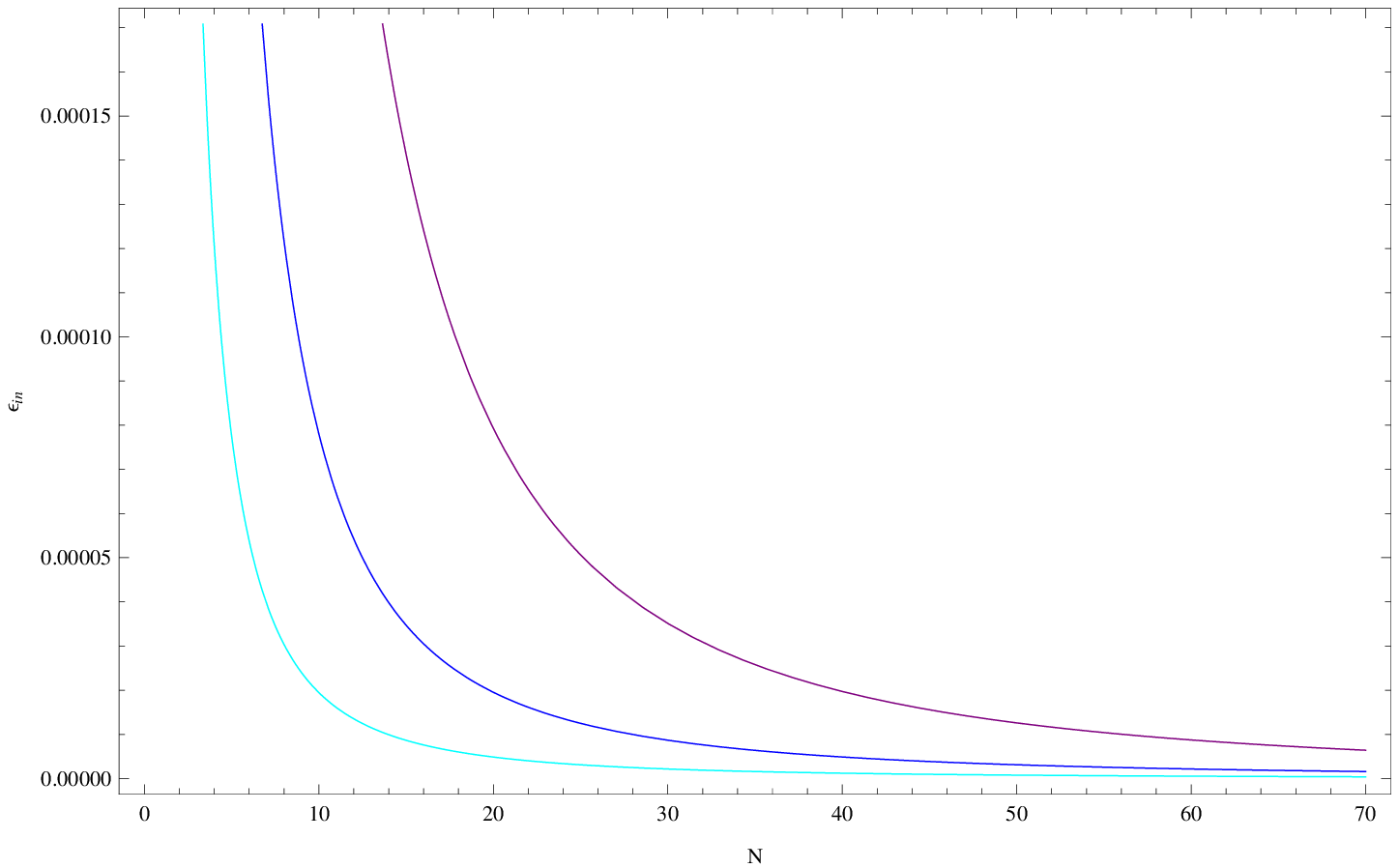}
    \includegraphics[width=18pc]{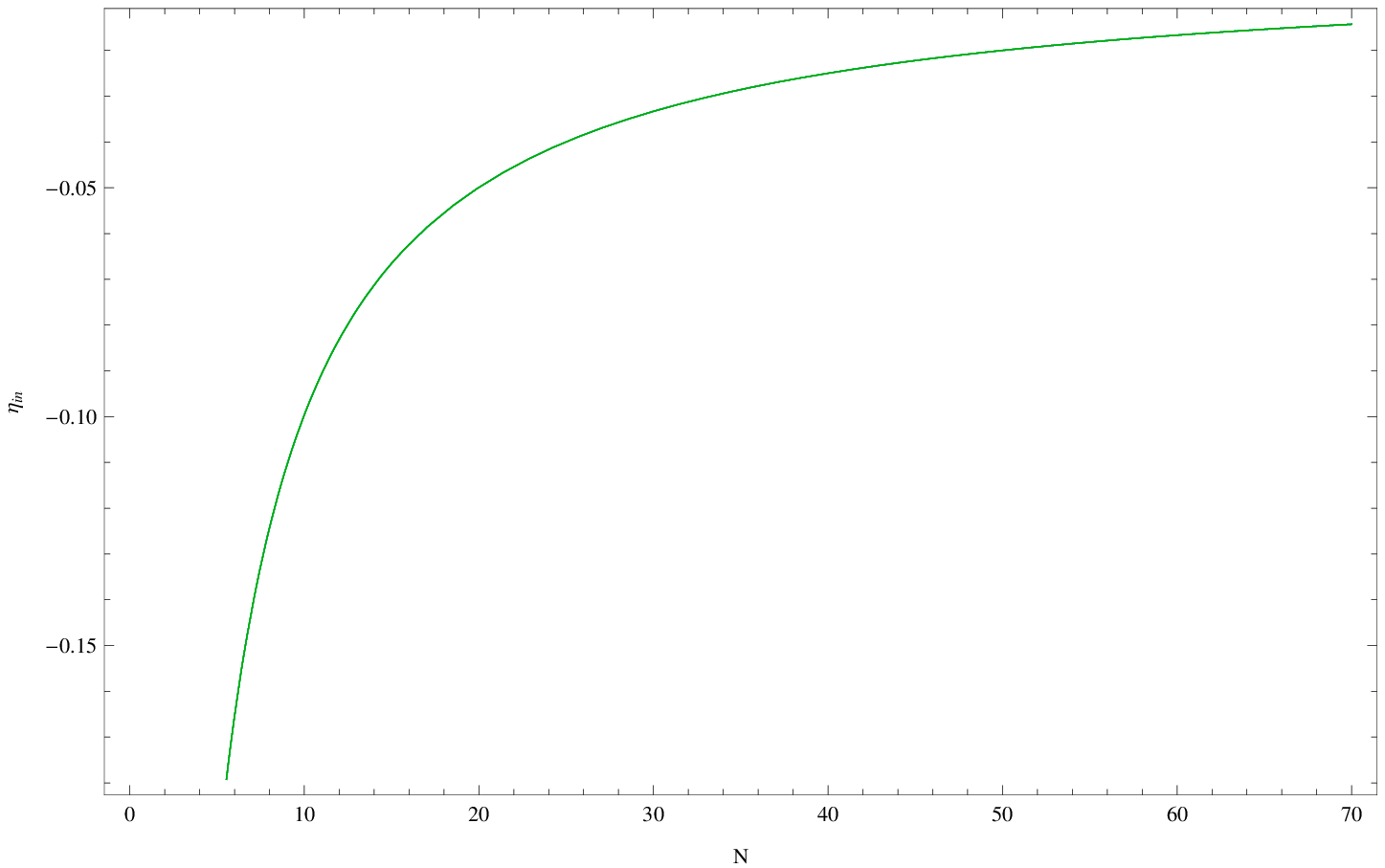}
\caption{The slow-roll indices with respect to the $e$-foldings number, $\epsilon_{in}$ on the left and $\eta_{in}$ on the right, for different values of $\alpha$ and $\beta$.}
\label{fig:SlowRollInd}
\end{figure}

\begin{figure}[h!] 
\centering
    \includegraphics[width=18pc]{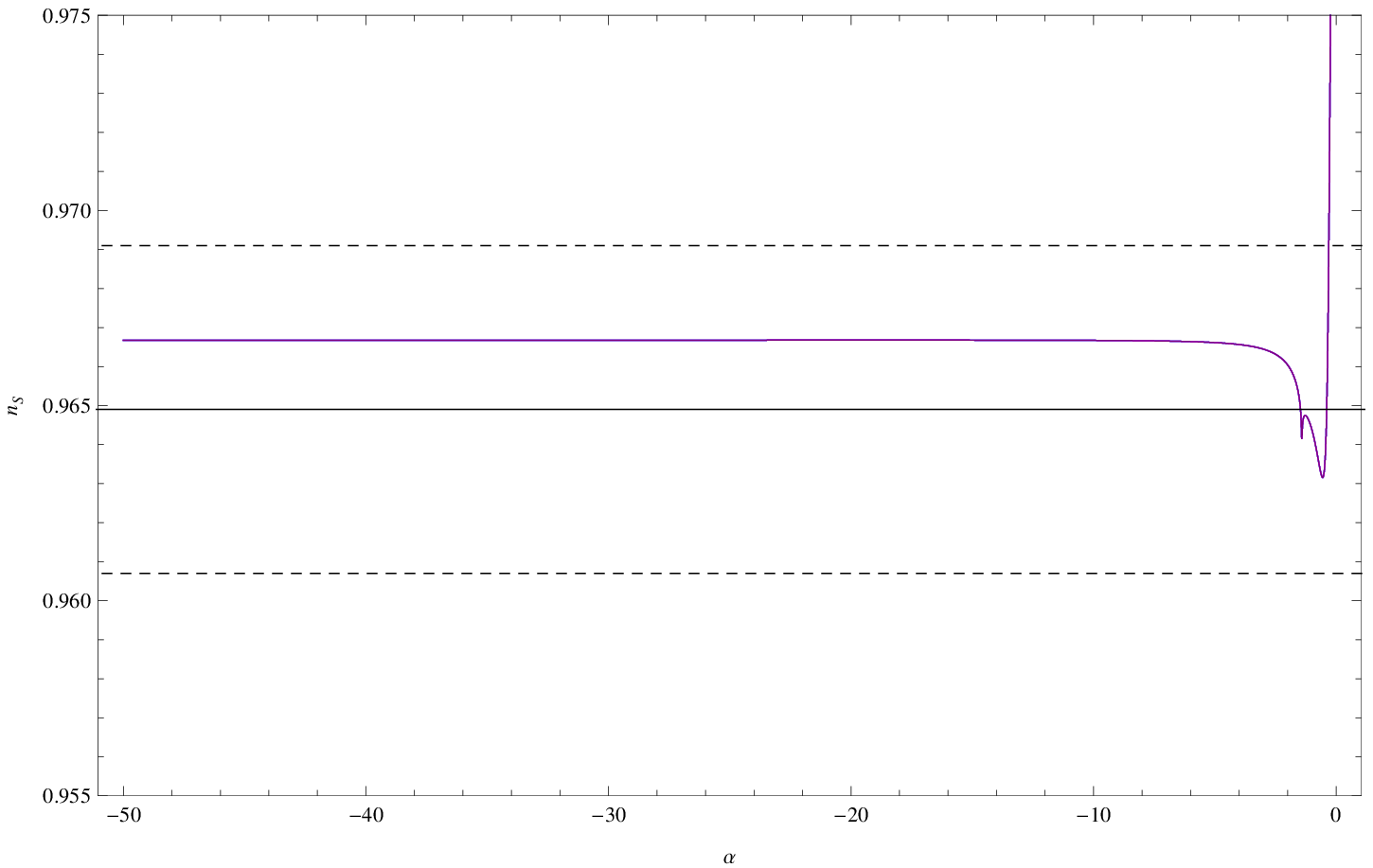}
    \includegraphics[width=18pc]{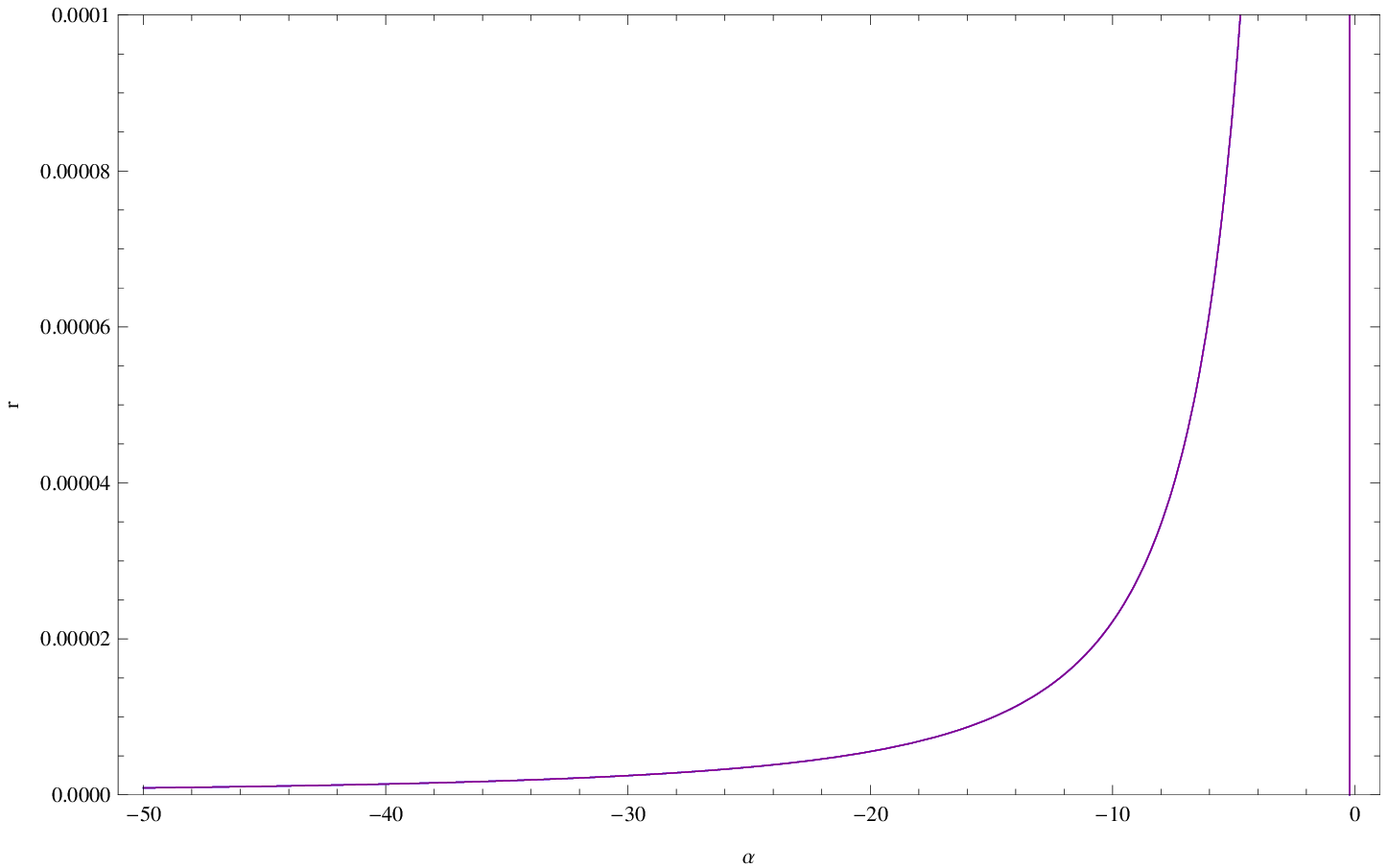}
\caption{The spectral index (on the left-hand-side) and the tensor-to-scalar ratio (on the right-hand-side) with respect to the $\alpha$ (only negative values), for $\kappa = 3.90584 \times 10^{-28}$eV$^{-1}$, $N = 60$ and arbitrary positive values for $\beta$. The horizontal line on the left plot stands for the $n_{S} = 0.9649$ obtained from the Planck collaboration, while the dashed horizontal lines denote the deviations within $95 \%$ significance.}
\label{fig:Observe1}
\end{figure}

\begin{figure}[h!] 
\centering
    \includegraphics[width=28pc]{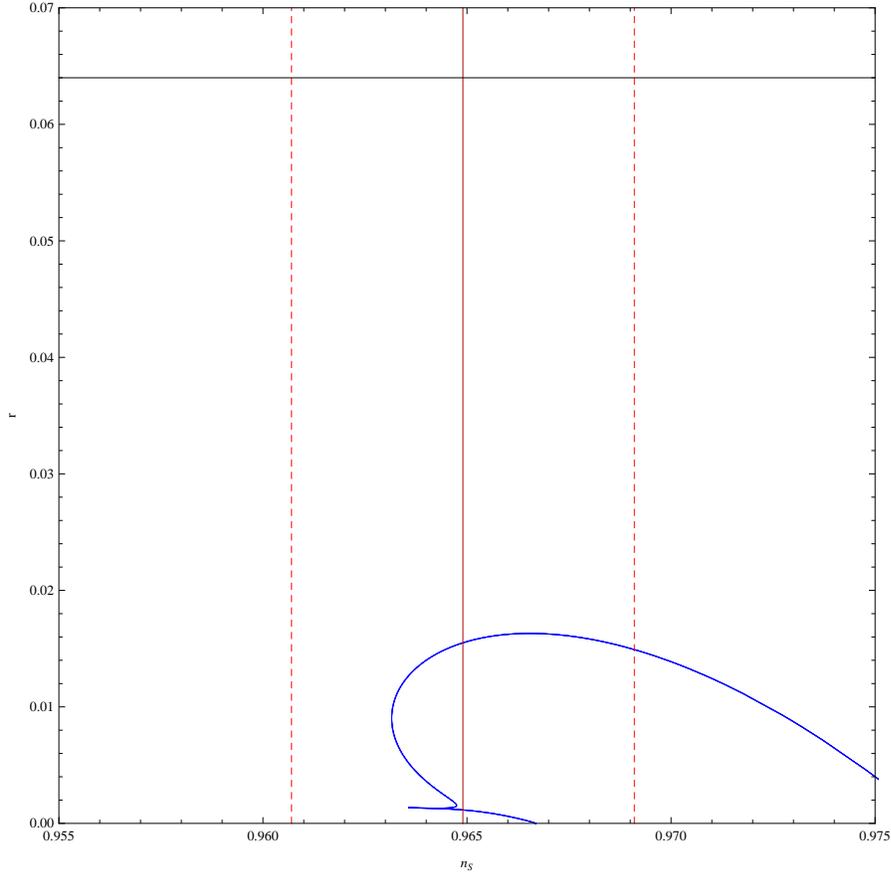}
\caption{The tensor-to-scalar ratio versus the spectral index for $\kappa = 3.90584 \times 10^{-28}$eV$^{-1}$, $N = 60$ and arbitrary positive values for $\beta$, as $\alpha$ varies between $-200$ and $0$eV. The horizontal black line stands for the upper limit $r = 0.064$ given by the Planck collaboration, the vertical crimson line denotes $n_{S} = 0.9649$, while the dashed vertical red lines denote its deviations within $95 \%$ significance, also given by the Planck collaboration.}
\label{fig:Observe2}
\end{figure}

We shall express all the physical quantities in terms of the $e$-foldings number, defined as follows,
\begin{equation}\label{eq:e-folds}
N = \int_{t_{in}}^{t_{fin}} H \mathrm{d}t \, ,
\end{equation}
where $t_{in}$ and $t_{fin}$ the initial and final moments of inflation. If the slow-roll approximation holds true, we easily see that,
\begin{equation}
N = \int_{t_{in}}^{t_{fin}} H \mathrm{d}t = \int_{\phi_{in}}^{\phi_{fin}} \dfrac{H}{\dot{\phi}} \mathrm{d}\phi \simeq \int_{\phi_{fin}}^{\phi_{in}} \dfrac{V(\phi)}{ V'(\phi) } \mathrm{d}\phi \, ;
\end{equation}
substituting $V(\phi)$ and its derivative from Eq. (\ref{eq:WoodsSaxon}), we can easily calculate the $e$-foldings number as a function of the scalar field in the beginning (and during) inflation,
\begin{equation}
N = \dfrac{\beta \left( 2\alpha\kappa \phi_{in} + \sqrt{2} \alpha + 2 \right) - 2\beta \ln \left( -\frac{\beta}{2} \left(\sqrt{2}\alpha + 2\right) \right) + 2 e^{\alpha\kappa \phi_{in}}}{2 \alpha^2 \beta} \, .
\end{equation}
Solving this with respect to the value of the scalar field in the beginning of inflation we have
\begin{equation}\label{eq:InflationField}
\phi_{in} = -\frac{-2 \ln \left(-\frac{\beta}{2} \left( \sqrt{2}\alpha + 2 \right) \right)+\sqrt{2} \alpha +2 \mathrm{W}\left( -\frac{1}{2} \left( \sqrt{2}\alpha + 2 \right) e^{-\frac{\alpha}{\sqrt{2}} + \alpha^2 N - 1} \right)-2 \alpha^2 N + 2}{2 \alpha\kappa} \, ,
\end{equation}
where $\mathrm{W}(x)$ is the Lambert function (also knows as the ``product logarithm''), that offers the inverse of $f(z) = z e^{z}$ for any complex number $z$. We shall assume that the initial value of the scalar field is the one corresponding to horizon crossing during inflation. The value of the scalar field in Eq. (\ref{eq:InflationField}) is given with respect to the $e$-foldings number, hence it may be said to represent the evolution of the scalar field over time (over $N$) during inflation. Substituting to the slow-roll indices of Eq. (\ref{eq:SlowRoll}), we may obtain these with respect to the $N$.
\begin{align}
\epsilon_{in} (N) &= \dfrac{2 \alpha ^2 \exp \left(\sqrt{2} \alpha +2 \mathrm{W}\left(-\frac{1}{2} \left(\sqrt{2} \alpha +2\right) e^{-\frac{\alpha }{\sqrt{2}}+\alpha^2 N - 1}\right)+2\right)}{\left(\left(\sqrt{2} \alpha +2\right) e^{\alpha^2 N}-2 \exp \left(\frac{\alpha }{\sqrt{2}}+\mathrm{W}\left(-\frac{1}{2} \left(\sqrt{2} \alpha +2\right) e^{-\frac{\alpha }{\sqrt{2}}+\alpha^2 N - 1}\right)+1\right)\right)^2} \, , \\
\eta_{in} (N) &= -\dfrac{\alpha ^2 \beta  \left(-\beta -\frac{1}{2} \left(\sqrt{2} \alpha +2\right) \beta  \exp \left(-\frac{\alpha }{\sqrt{2}}-\mathrm{W}\left(-\frac{1}{2} \left(\sqrt{2} \alpha +2\right) e^{-\frac{\alpha }{\sqrt{2}}+\alpha^2 N - 1}\right)+\alpha^2 N - 1\right)\right)}{\left(\beta -\frac{1}{2} \left(\sqrt{2} \alpha +2\right) \beta  \exp \left(-\frac{\alpha }{\sqrt{2}}-\mathrm{W}\left(-\frac{1}{2} \left(\sqrt{2} \alpha +2\right) e^{-\frac{\alpha }{\sqrt{2}}+\alpha^2 N-1}\right)+\alpha^2 N - 1\right)\right)^2}
\end{align}
As we observe in Fig. \ref{fig:SlowRollInd}, both of them converge very fast to small -almost zero- values. Also $\eta_{in}$ does not seem to depend significantly on parameters $\alpha$ and $\beta$, despite appearing on its expression, while $\epsilon_{in}$ tends faster to zero for lower values of $\alpha$. We need to note that that we used only negative values of $\alpha$, as any $\alpha > -\dfrac{\sqrt{2}}{2}$, that stood for $\phi^{*} = \phi_{fin}$, gave infinitely increasing $\epsilon_{in}$.
Using Eqs. (\ref{eq:SpectralInd}) and (\ref{eq:TensorScalarRatio}), we can eventually calculate the spectral index, which is,
\begin{equation}\label{eq:ns}
n_{S} = 1 - \dfrac{\alpha ^2 \left(2 \mathrm{W}\left(-\frac{1}{2} \left(\sqrt{2} \alpha +2\right) e^{-\frac{\alpha }{\sqrt{2}}+\alpha^2 N - 1}\right)+1\right)}{\left(\mathrm{W}\left(-\frac{1}{2} \left(\sqrt{2} \alpha +2\right) e^{-\frac{\alpha }{\sqrt{2}}+\alpha^2 N - 1}\right)+1\right)^2} \, ,
\end{equation}
and the tensor-to-scalar ratio,
\begin{equation}\label{eq:r}
r = \dfrac{8 \alpha ^2}{\left(\mathrm{W}\left(-\frac{1}{2} \left(\sqrt{2} \alpha +2\right) e^{-\frac{\alpha }{\sqrt{2}}+\alpha^2 N - 1}\right)+1\right)^2} \, ,
\end{equation}
It is easy, in this form, to compare them with the observational constraint coming from the latest Planck data. As it can be see in Fig. \ref{fig:Observe1}, the only parameter playing some important role in the size and value of the spectral index and especially of the tensor-to-scalar ratio, is $\alpha$. However, as long as $| \alpha |$ is sufficiently large, no problem arises, as $n_{S} \simeq 0.966676$ and $r \ll 0.064$. Also, the value of $V_0$ does not affect at all the observational indices. However, as we demonstrate in the next section, $V_0$ affects strongly the reheating era.
In Fig. \ref{fig:Observe2}, we plot the parametric plot of $n_{S}$ and $r$ for our model, given $\kappa = 3.90584 \times 10^{-28}$eV$^{-1}$, $N = 60$, $\beta > 0$ and a varying $\alpha < 0$. The blue curve corresponds to their ``trajectory'' with respect to $\alpha$. The horizontal black line stands for the upper limit $r = 0.064$ given by the Planck collaboration, the vertical crimson line denotes $n_{S} = 0.9649$, while the dashed vertical red lines denote its deviations within $95 \%$ significance, also given by the Planck collaboration. We can plainly see that the two observable quantities are within the observationally allowed range of values.
Having discussed the inflationary phenomenology of the model, and upon verifying its viability, what remains now is to prove that it can also provide a viable reheating era. This is the subject of the next section.

\begin{figure}[h!] 
\centering
    \includegraphics[width=18pc]{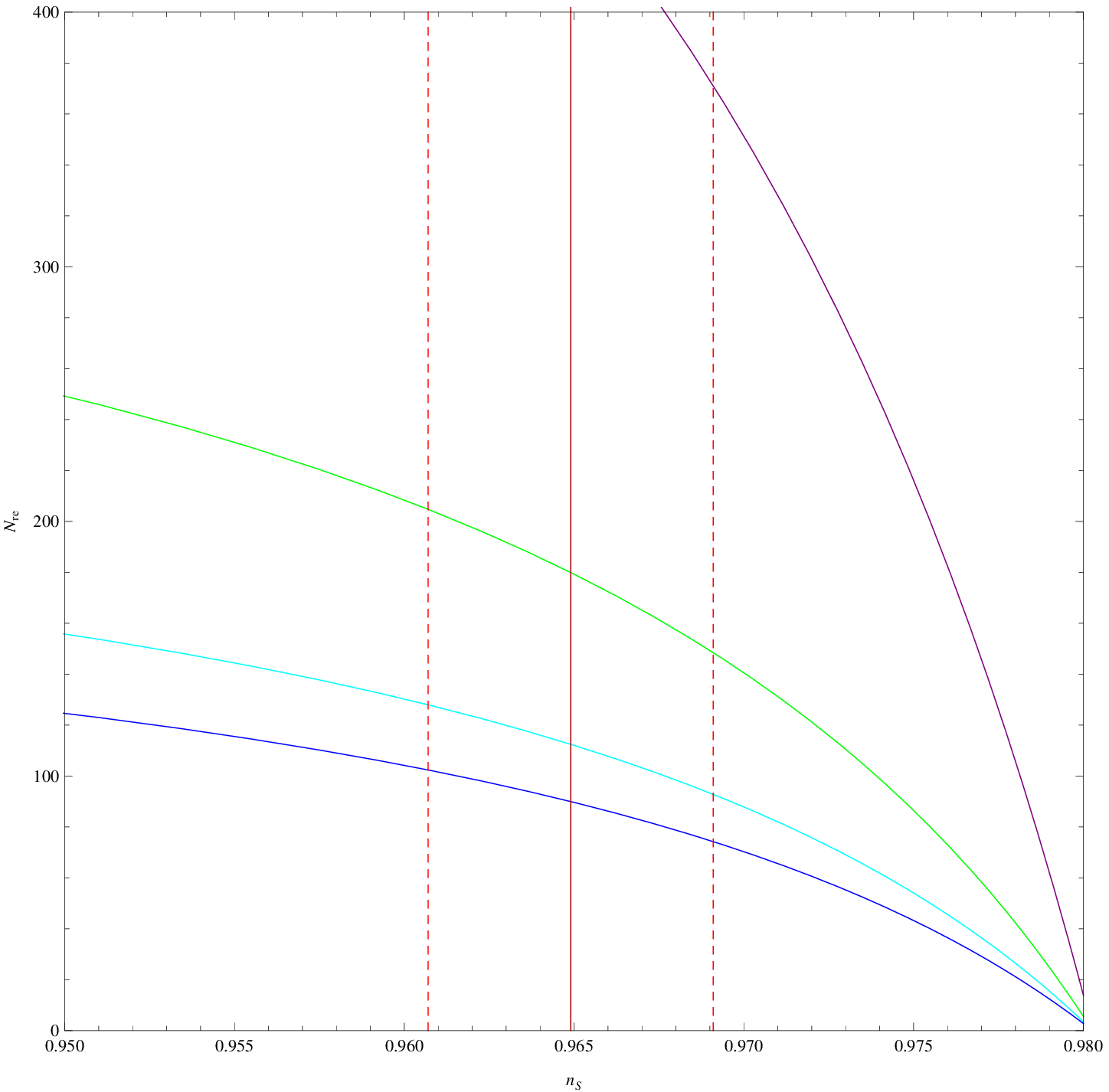}
    \includegraphics[width=18pc]{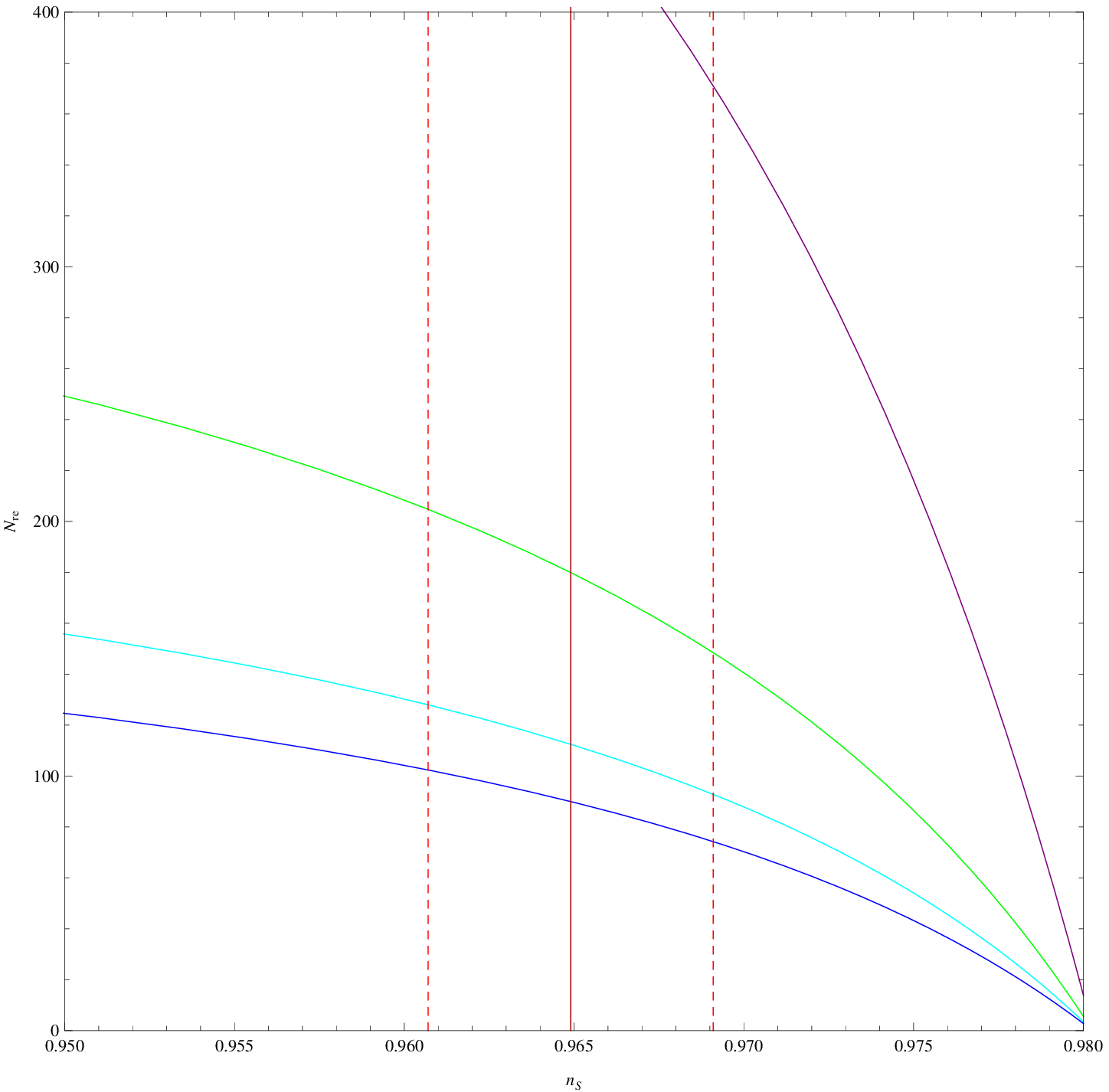}
    \includegraphics[width=18pc]{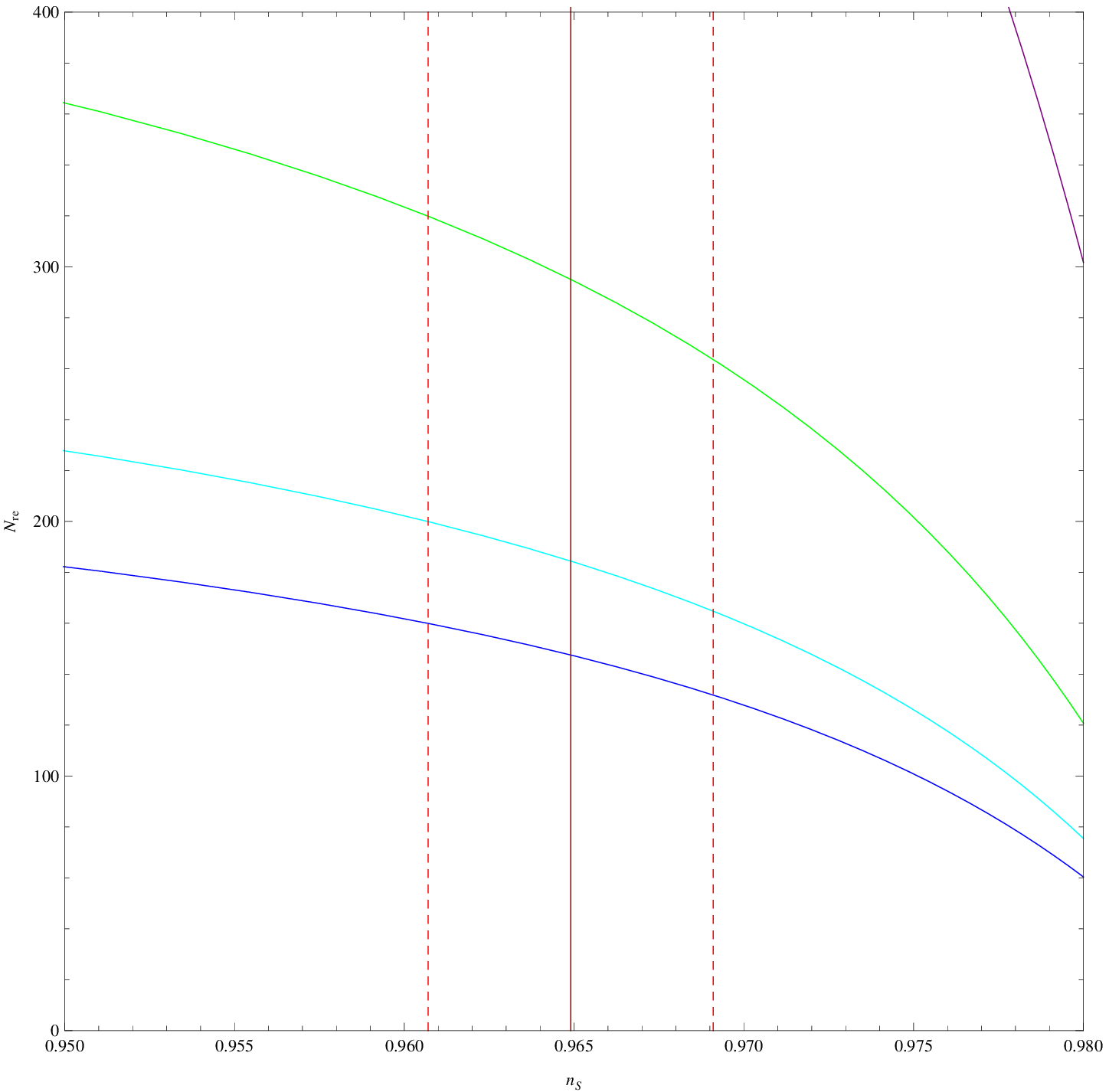}
\caption{The duration of the reheating $N_k$ versus the spectral index for $\kappa = 3.90584 \times 10^{-28}$eV$^{-1}$, $A_{s} = 1.90461 10^{-9}$, and $T_{0}=3.19 \times 10^{-5}$eV. Also the parameters $\alpha$ and $\beta$ are given the same values of the previous section that make simultaneously the inflationary theory viable. $w_{re}$ is equal to $-\dfrac{1}{3}$ (blue curves), $-\dfrac{1}{5}$ (cyan curves), $0$ (light green curves) and $\dfrac{1}{5}$ (purple curves). The vertical crimson line denotes $n_{S} = 0.9649$, while the dashed vertical red lines denote its deviations within $95 \%$ significance, also given by the Planck collaboration. The upper left plot corresponds to $V_0=10^{2}$eV$^{4}$, the upper right to $V_0=10^{-10}$eV$^{4}$ and the bottom plot to $V_0=10^{-60}$eV$^{4}$. The reheating temperature is crucially affected by the parameter $V_0$.}
\label{fig:NreNs}
\end{figure}

\section{The Reheating Era}

As we demonstrated in the previous section, the inflationary era generated by the Woods-Saxon scalar field model is quite well fitting the observational data, deeming this a good candidate for the description of early-time dynamics. However, as has been stressed in the literature, the graceful exit from inflation is not enough, since the temperature of the Universe has decreased significantly and in effect it must be somehow be increased by some internal mechanism in the post-inflationary era, so that cosmic evolution will continue as we know from the standard Big Bang model \cite{Kofman:1997yn,Albrecht:1982mp, Shtanov:1994ce, Kofman:1994rk, Linde:2005ht, Rinaldi:2015uvu,Amin:2014eta}. This problem is usually addressed by the very own structure of the inflationary potential, that is expected to reach either a minimum, or some times an inflection point, where the inflation-generating scalar field will also generate the reheating of the Universe \cite{Amin:2014eta, Martin:2014nya, Gong:2015qha, Cai:2015soa, Cook:2015vqa, Rehagen:2015zma, deFreitas:2015xxa, Rinaldi:2015uvu, Amin:2015lnh, Choi:2016eif, deHaro:2016hsh, Ueno:2016dim, Eshaghi:2016kne, Maity:2016uyn, Tambalo:2016eqr, Fei:2017fub, deHaro:2017nui, Oikonomou:2017bjx, Maity:2017thw, Artymowski:2017pua, German:2018wrx, Ji:2019gfy, Dimopoulos:2019gpz}. Of course, this idea has not come without restraints, since classical scalar fields, like the Higgs, are proved to block or delay the reheating \cite{Freese:2017ace, Fonseca:2018xzp}, with some references claiming that auxiliary scalar fields \cite{Dimopoulos:2018wfg} or some non-minimal gravitational couplings \cite{Koh:2018qcy} are essential to the reheating phase. An alternative way to address this problem is by seeking a safe transition from the inflationary potential decay to the radiation-dominated era afterwards \cite{Turzynski:2018zup}.

Our scope here is merely to test the Woods-Saxon scalar field towards its capability to generate the post-inflationary reheating. In doing so, we need to calculate the thermalization temperature of the Universe, $T_{re}$, and the duration of the reheating, $N_{re}$ (again in terms of the $e$-foldings number), as well as to provide an effective equation of state for the post-inflationary matter fields, which should give a barotropic index greater than $-\dfrac{1}{3}$, so that acceleration ceases, and lower than $1$, so that causality is preserved. In addition, the spectral index, $n_{S}$, must be maintained in the value it obtains at the end of inflation, so that the scalar perturbation modes maintain their amplitude during the radiation-dominated era and in the Cosmic Microwave Background. More specifically, we define the duration, $N_{re}$, as the number of $e$-foldings from the final moment of inflation, $N_{fin} \sim 50-60$, till the equation of state transits to $w_{eff} = \dfrac{1}{3}$ (the radiation-dominated era) and the thermalization temperature, $T_{re}$, is the temperature reached at this the exact time instance that the Universe enters the radiation domination era, although this definition has a flaw, since it is not consistent with $w_{eff} = \dfrac{1}{3}$ during reheating, see Ref. \cite{Cook:2015vqa} for discussion.

In the analysis that follows we shall use the notation and terminology of Refs. \cite{Martin:2014nya, Gong:2015qha, Cai:2015soa, Cook:2015vqa, Rehagen:2015zma, deFreitas:2015xxa}. Naming $a_{fin}$ and $\rho_{fin}$ the scale factor and the energy density at the end of inflation and, similarly, $a_{re}$ and $\rho_{re}$ at the end of the reheating, we may easily apply the solution of Friedmann's equation Eq. \ref{eq:Friedmann} as,
\begin{equation}
\dfrac{\rho_{fin}}{\rho_{re}} = \left( \dfrac{a_{fin}}{a_{re}} \right)^{-3(1+w_{re})} \, ,
\end{equation}
where $w_{re}$ the barotropic index at the end of reheating. Rewriting this with respect to $e$-foldings number, it is easy to calculate the duration of the reheating, which is \cite{Cook:2015vqa},
\begin{equation}\label{eq:Duration1}
N_{re} = \dfrac{1}{3(1+w_{re})} \ln \left( \dfrac{\rho_{fin}}{\rho_{re}} \right) = \dfrac{1}{3(1+w_{re})} \ln \left( \dfrac{45 V_{fin}}{\pi^2 g_{re} T_{re}^4} \right) \, ,
\end{equation}
since $\rho_{fin} = \dfrac{3}{2} V_{fin}$ and $\rho_{re} = \dfrac{\pi^2}{30} g_{re} T_{re}^4$, where $V_{fin}$ is the value of the potential at the end of inflation and $g_{re}$ is the number of relativistic species during the reheating era, which is estimated by \cite{Akrami:2018odb} to be $\sim 100$. In the same manner, the temperature at the end of the reheating is \cite{Cook:2015vqa},
\begin{equation}
T_{re} = T_{0} \left( \dfrac{a_{0}}{a_{re}} \right) \left( \dfrac{43}{11 g_{re}} \right)^{\frac{1}{3}} = T_{0} \left( \dfrac{a_{0}}{a_{eq}} \right) e^{N_{RD}} \left( \dfrac{43}{11 g_{re}} \right)^{\frac{1}{3}} \, ,
\end{equation}
where $a_{eq}$ the scale factor in the matter-radiation equality moment and $N_{RD}$ the duration (in $e$-folds) of the radiation-dominated era, since $a_{re} = a_{eq} e^{N_{RD}}$, and $T_0$ is the present day temperature $T_0\sim 3.19\times 10^{-5}$eV. We can further write $T_{re}$ \cite{Cook:2015vqa},
\begin{equation}\label{eq:Temperature1}
T_{re} = a_{0} T_{0} \dfrac{H_{k}}{k} e^{-N} e^{-N_{re}} \left( \dfrac{43}{11 g_{re}} \right)^{\frac{1}{3}} \, ,
\end{equation}
where $N_{k}$ and $H_{k}$ is the duration and the Hubble rate of the pivot scale $k = a_{k} H_{k}$ for some observation, crossing outside the Hubble radius and starting at the end of inflation. We shall use the result of the Planck collaboration \cite{Akrami:2018odb}, which is $\dfrac{k}{a_{k}} \simeq 0.05 \; Mpc^{-1}$.

For $w_{eff} \neq \dfrac{1}{3}$, combining Eqs. (\ref{eq:Duration1}) and (\ref{eq:Temperature1}), we may easily find that the duration of reheating is given as,
\begin{equation}\label{eq:Duration2}
N_{re} = \dfrac{4}{1 - 3 w_{re}} \left( 61.6 - N_{k} - \ln \left( \dfrac{V_{fin}^{\frac{1}{4}}}{H_{k}} \right) \right) \, ,
\end{equation}
while the temperature is \cite{Cook:2015vqa},
\begin{equation}\label{eq:Temperature2}
T_{re} = \left[ \left( \dfrac{43}{11 g_{re}} \right)^{\frac{1}{3}} \dfrac{T_{0} H_{k}}{0.05} e^{-N_{k}} \right]^{\frac{3(1+w_{re})}{3 w_{re} - 1}} \left[ \dfrac{45 V_{fin}}{\pi^2 g_{re}} \right]^{1 - 3 w_{re}} \, ,
\end{equation}
where $H_{k}$ and $N_{k}$ are given as a consequence of a specific model, namely of a specific scalar field.

We know that the duration in $e$-foldings is given by
\begin{equation*}
N_{k} = \int_{t_{k}}^{t_{fin}} H(t) \mathrm{d}t = \kappa \int_{\phi_{fin}}^{\phi_{k}} \dfrac{V(\phi)}{ \frac{\mathrm{d}V}{\mathrm{d}\phi}} \mathrm{d} \phi \, .
\end{equation*}
Given that the scalar field follows a Woods-Saxon potential from Eq. (\ref{eq:WoodsSaxon}), the integral yields
\begin{equation}\label{eq:Nk}
N_{k} = \dfrac{\beta  \left(2 \alpha\kappa \phi_{k} + \sqrt{2}\alpha + 2 \right) - 2\beta \ln \left(-\frac{\beta}{2} \left( \sqrt{2}\alpha + 2 \right) \right) + 2 e^{\alpha\kappa  \phi_{k}}}{2\alpha^2 \beta} \, .
\end{equation}
From this, we may calculate the value of the scalar field in the pivot scale, $\phi_{k}$, and thus the value of the potential,
\begin{equation}\label{eq:Vend}
V_{fin} = \dfrac{V_{0}}{\beta \exp \left(\frac{1}{2} \left(-2 \ln \left(-\frac{\beta}{2} \left( \sqrt{2}\alpha + 2 \right) \right) + \sqrt{2}\alpha - 2\alpha^2 N_{k} + 2 \mathrm{W}\left(-\frac{1}{2} \left( \sqrt{2}\alpha + 2 \right) e^{-\frac{\alpha}{\sqrt{2}} + \alpha^2 N_{k} - 1} \right)+ 2 \right)\right) + 1} \, ,
\end{equation}
the slow-roll indices, $\epsilon_{k}$ and $\eta_{k}$, and finally the Hubble rate for the pivot scale,
\begin{equation}\label{eq:Hk}
H_{k} = \pi \sqrt{ \dfrac{8 A_{s} \epsilon_{k}}{\kappa} } = \dfrac{4\pi}{\sqrt{\kappa}} \sqrt{ \dfrac{ \alpha^2 A_{s} \exp \left( \sqrt{2}\alpha + 2 \mathrm{W}\left(-\frac{1}{2} \left(\sqrt{2} \alpha + 2 \right) e^{-\frac{\alpha }{\sqrt{2}} + \alpha^2 N_{k} - 1}\right) + 2 \right)}{\left(\left( \sqrt{2}\alpha + 2 \right) e^{\alpha^2 N_{k}} - 2 \exp \left(\frac{\alpha}{\sqrt{2}} + \mathrm{W}\left(-\frac{1}{2} \left(\sqrt{2}\alpha +2\right) e^{-\frac{\alpha}{\sqrt{2}} + \alpha^2 N_{k} - 1}\right) + 1 \right)\right)^2}} \, ,
\end{equation}
where $A_{s}$ the amplitude of curvature perturbations, which the latest Planck data \cite{Akrami:2018odb} suggest that it is $A_{s} = 1.90461 \times 10^{-9}$, which we shall use.
\begin{figure}[h!] 
\centering
    \includegraphics[width=18pc]{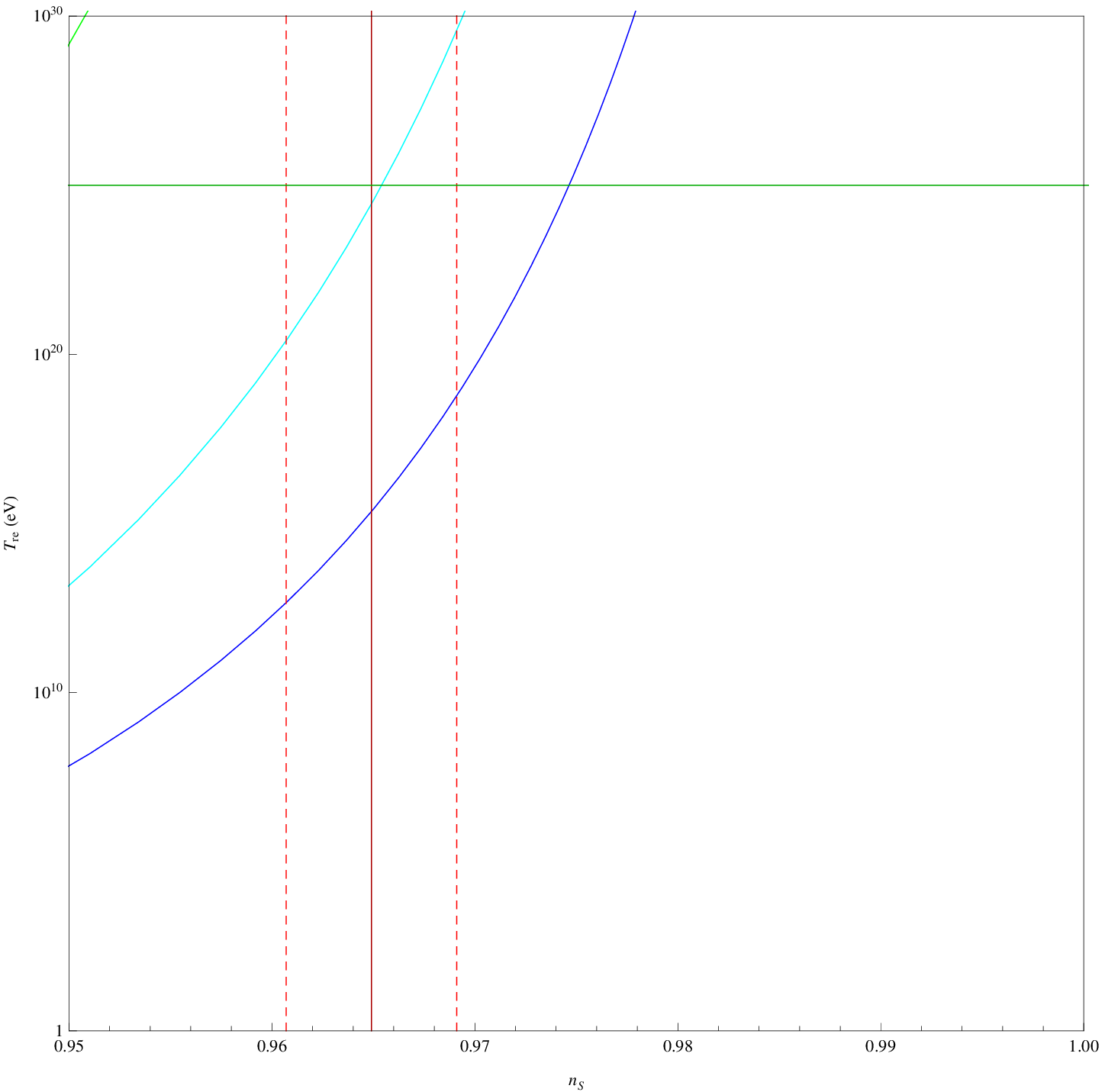}
    \includegraphics[width=18pc]{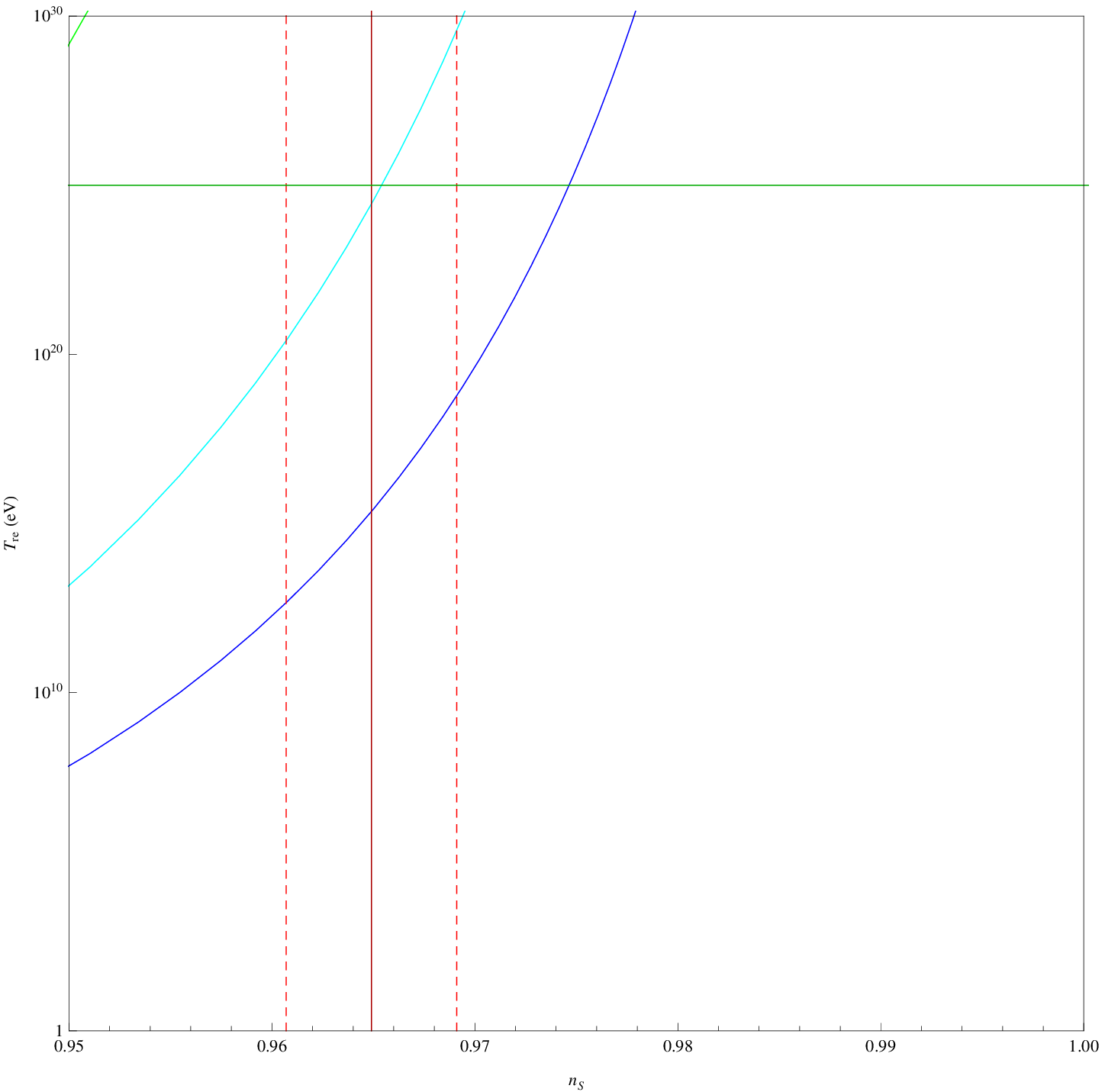}
    \includegraphics[width=18pc]{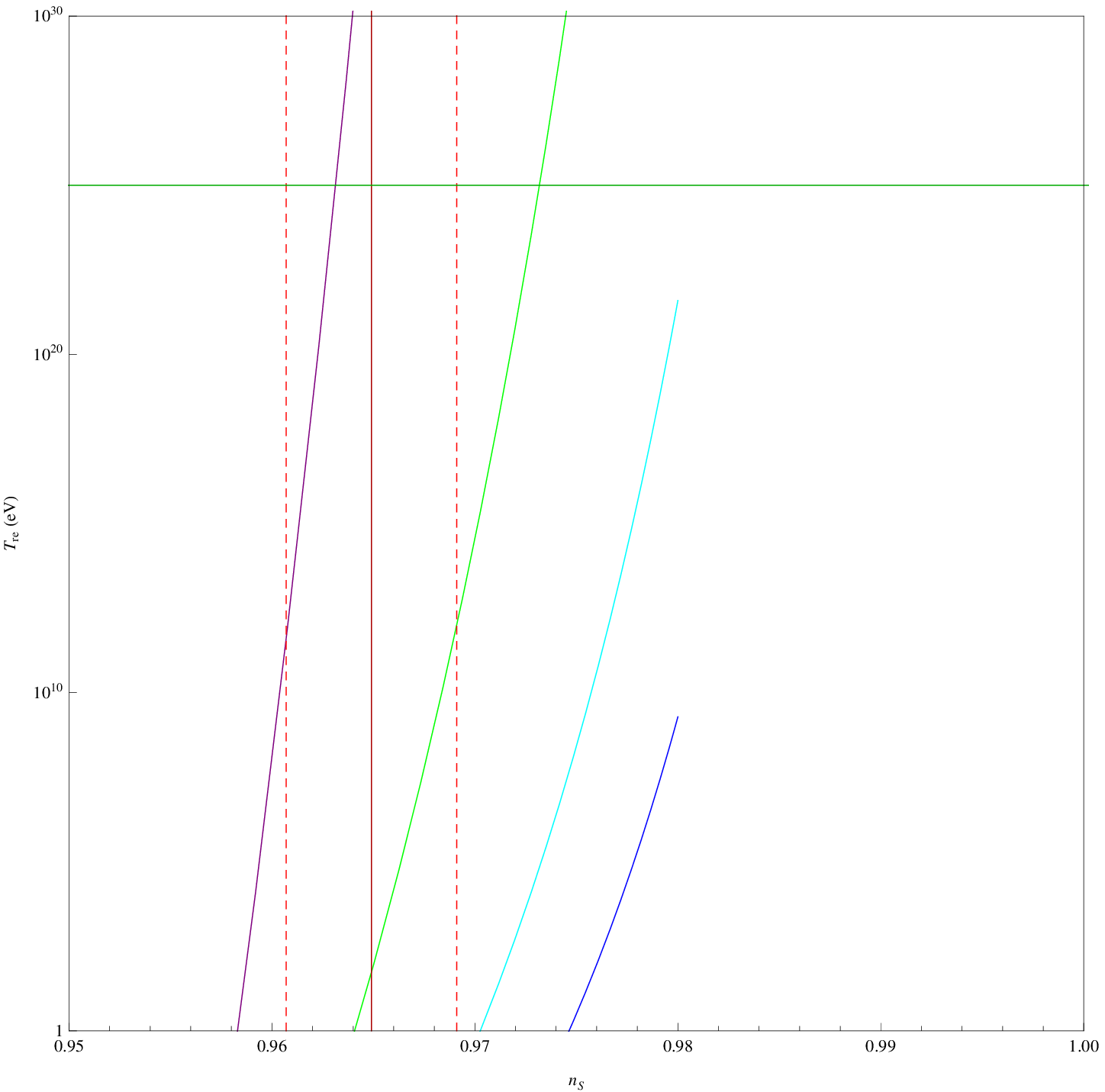}
\caption{The reheating temperature $T_{re}$ (eV) versus the spectral index for $\kappa = 3.90584 \times 10^{-28}$eV$^{-1}$, $A_{s} = 1.90461 \times 10^{-9}$ and $T_{0}=3.19 \times 10^{-5}$eV. Also the parameters $\alpha$ and $\beta$ are given the same values of the previous section that make simultaneously the inflationary theory viable. $w_{re}$ is equal to $-\dfrac{1}{3}$ (blue curves), $-\dfrac{1}{5}$ (cyan curves), $0$ (light green curves) and $\dfrac{1}{5}$ (purple curves). The vertical crimson line denotes $n_{S} = 0.9649$, while the dashed vertical red lines denote its deviations within $95 \%$ significance, also given by the Planck collaboration. The horizontal green curve denotes the upper bound on the reheating temperature imposed by the inflationary scale $T_i=10^{25}$eV. The upper left plot corresponds to $V_0=10^{2}$eV$^{4}$, the upper right to $V_0=10^{-10}$eV$^{4}$ and the bottom plot to $V_0=10^{-60}$eV$^{4}$. The reheating temperature is crucially affected by the parameter $V_0$.}
\label{fig:TreNs}
\end{figure}
Substituting Eqs. (\ref{eq:Nk}), (\ref{eq:Vend}) and (\ref{eq:Hk}) in Eq. (\ref{eq:Duration2}) and in Eq. (\ref{eq:Temperature2}), we may calculate the duration and the temperature of the reheating respectively for different barotropic indices, $w_{re}$. Generally, we know that $w_{re} = -\dfrac{1}{3}$ means that we enter a constantly expanding Universe and thus exiting inflation, while $w_{re} > -\dfrac{1}{3}$ describes decelerating expanding Universe. We also know that $w_{re} = \dfrac{1}{3}$ indicates the radiation-dominated era, that has to follow the early stages of the Universe, while $w_{re} > 1$ corresponds to models not respecting the spacetime causality. At the same time, we have to ensure that the spectral index remains close to the values observed in the Cosmic Microwave Background radiation, so that the the curvature perturbations created throughout the inflationary era are preserved during the early Universe and until the radiation-dominated era, after which they shall form the observed structures.

In order to check the aforementioned constraints for the model at hand, we plot both the duration $N_{re}$ and the temperature $T_{re}$ (in eV) of the reheating with respect to the spectral index $n_{S}$, as the duration of the pivot scale $N_{k}$ varies, for a wide selection of matter content with barotropic indices in the interval $\Big[ -\dfrac{1}{3} , \dfrac{1}{3} \Big)$. These are presented in Figs. \ref{fig:NreNs} and \ref{fig:TreNs} respectively. The blue curves represent $w_{re} = -\dfrac{1}{3}$, cyan curves represent $w_{re} = -\dfrac{1}{5}$, light green curves represent $w_{re} = 0$ and purple curves $w_{re} = \dfrac{1}{5}$, as different possible stages in the transition. Similarly, the vertical crimson line stands for $n_{S} = 0.9649$, while the dashed vertical red lines for its deviations within $95 \%$ significance, also given by the Planck collaboration. Also the horizontal green curve represents the upper limit on the reheating temperature, coming from the inflationary scale $T_i\sim 10^{25}$eV. The upper left plot of Fig. \ref{fig:TreNs} corresponds to $V_0=10^{2}$eV$^{4}$, the upper right to $V_0=10^{-10}$eV$^{4}$ and the bottom plot to $V_0=10^{-60}$eV$^{4}$. Also the parameters $\alpha$ and $\beta$ are given the same values of the previous section that make simultaneously the inflationary theory viable.

As it can be seen in all plots, the reheating temperature is crucially affected by the parameter $V_0$, and in addition the reheating era is generated by the Woods-Saxon scalar theory in a viable way. However, for large values of the parameter $V_0$, which recall that controls the depth of the Woods-Saxon potential, several equations of state for the scalar field are excluded. For example in the case $V_0=10^{2}$eV$^{4}$, the equation of state parameter values $w_{re}=0$ and $w_{re}=1/5$ are excluded since these generate extremely high reheating temperatures. Finally, the instant reheating mechanism for the model, which corresponds to $N_k=0$, is excluded for all the values of the equation of state parameter and for all the values of $V_0$.

\section{Conclusions}

The focus in this work was on the realization of an inflationary scenario in the context of a canonical scalar field with Woods-Saxon potential, along with its capability of producing a viable subsequent reheating era. As we demonstrated, both the inflationary era and the reheating era can be successfully described by the Woods-Saxon scalar field model. Specifically, the early-time accelerating expansion is naturally driven by the slow-rolling of the scalar field and the theoretical predictions about the scalar and tensor perturbations are successfully compared to the observational data of the Planck collaboration. In addition, we showed that a reheating era maybe produced by the Woods-Saxon scalar theory, in which the maximum reheating temperature is within the existing constraints, for the same values of the free parameters that make the inflationary theory viable.

Thus the same set of the free parameters values can produce a viable inflationary era and at the same time a viable reheating era. Also it seems that the parameter $V_0$ of the Woods-Saxon potential crucially affects the reheating era, while it does not affect at all the inflationary era. This parameter determines the depth of the potential. Another interesting feature of our work is that the value of the scalar field for which the graceful exit occurs, coincides with the inflection point of the scalar potential, a feature which can be interesting phenomenologically. Several authors support the idea of an inflection point in the potential generating reheating \cite{Choi:2016eif, Ueno:2016dim, Germani:2017bcs, Fei:2017fub,German:2018wrx}.

Finally, let us note that, inflation might not be the result of a single slow-rolling scalar field, but maybe it is generated by many scalar fields, hence the reheating afterwards too can also be the results of many scalar fields \cite{Assadullahi:2016gkk, Abedi:2016sks, Iyer:2017qzw, Achucarro:2017ing, Carrilho:2018ffi, Christodoulidis:2018qdw,DeCross:2016cbs,DeCross:2016fdz,DeCross:2015uza,Schutz:2013fua,Kaiser:2013sna, Dimopoulos:2018wfg, Rudelius:2018yqi, McAneny:2019epy}. In this way, we may also consider a specific scalar field to dominate during inflation and another to dominate later, or more than one dominating at the same time.

\section*{Acknowledgments}

This work is supported by the DAAD program "Hochschulpartnerschaften mit Griechenland 2016" (Projekt
57340132) (V.K.O). V.K.O is indebted to Prof. K. Kokkotas for his hospitality in the IAAT, University of T\"{u}bingen.

\end{document}